\documentclass[useAMS,usenatbib]{mn2e}

\usepackage{graphicx}  
\usepackage{amsmath}
\usepackage{soul}
\usepackage{xcolor}
\setulcolor{red}
\usepackage{hyperref}
\usepackage{color}
\hypersetup{colorlinks,
 citecolor=blue,
 linkcolor=blue}

\def\aj{AJ}
\def\apj{ApJ}
\def\apjl{ApJ}
\def\apjs{ApJS}
\def\aap{A\&A}
\def\mnras{MNRAS}
\def\pasp{PASP}%
\def\araa{ARA\&A}

\title[Origin of Compact galaxies]{SDSS J122958.84+000138.0: A Compact, Optically red galaxy \thanks{Based on observations collected at the European Organisation for Astronomical Research in the Southern Hemisphere, Chile (programme 087.B-0841(A))}
}
   
\author[S. Paudel et al]{Sanjaya Paudel$^{1}$\thanks{E-mail:
sjy@kasi.re.kr}, Thorsten Lisker$^{2}$, Avon P. Huxor$^{2}$ and Chang H. Ree$^{1}$\\
$^{1}$Korea Astronomy and Space Science Institute, Daejeon 305-348, Republic of Korea\\
$^{2}$Astronomisches Rechen-Institut, Zentrum f\"ur Astronomie der Universit\"at Heidelberg, M\"onchhofstra\ss e 12-14, 69120 Heidelberg,
 Germany}
\begin{document}

\date{\today}

\maketitle

\label{firstpage}

\begin{abstract}

We report a new compact galaxy, SDSS J122958.84+000138.0 (SDSS J1229+0001), which has unique morphological and stellar population properties that are rare in observations of the nearby universe. SDSS J1229+0001 has an $r$-band absolute magnitude (M$_{r}$) and half-light radius (R$_{h}$) of $-$17.75  mag and  520 pc, respectively. Located in a fairly low density environment, morphologically it is akin to a typical early-type galaxy as it has a smooth appearance and red colour.  But, interestingly, it possesses centrally concentrated star forming activity with a significant amount of dust. We present  an analysis of structural and stellar population properties using archival images and VLT/FORS2 spectroscopy. Analysis of UKIDSS H-band image shows that the observed light distribution is better fitted with two components S\'ersic function with inner and outer component effective radii 190 and 330 pc, respectively. Whereas, overall half-light radius measured in H-band is much smaller compared to optical, i.e 290 pc. We prepared a Spectral Energy Distribution (SED) from optical to FIR and interpret it to derive star-formation rate, dust mass and stellar mass. We find that the SDSS J1229+0001 has dust mass M$_{dust}$ =  5.1 $\times$ 10$^{5}$ M$_{\sun}$ with a dust to stellar mass ratio log(M$_{dust}$/M$_{*}$) = $-$3.5. While the observed stellar population properties are -- to some extent -- similar to that of a typical S0 galaxy, a unified view from stellar population and structural properties may suggests that SDSS J1229+0001 is a {\it smoking gun} example of a compact early-type galaxy in formation.

\end{abstract}


\section{Introduction}

Correlations and anti-correlations between galaxy parameters, known as scaling relations, have been used to understand the underlying physical principles of galaxy formation.  Of particular interest is the size-luminosity relation, which has received considerable attention in recent studies of galaxy of formation and evolution (e.g. \cite{Kormendy85,Janz08}).  However, observational data exhibit  both a large scatter and a significant number of outliers \citep{Graham03,Kormendy09,Chen10,Misgeld11}. There exist a compact class of galaxies, compact ellipticals (cEs), which do not follow the Log(size)-magnitude relation and becomes an outlier \citep{Chilingarian09,Trujillo09,Paudel14}. They are compact, high surface brightness, metal-rich and possess a large velocity dispersion  compared to the majority of similar mass galaxies, i.e. early-type dwarfs (dEs). So, their origin and evolution are expected to be different from that of dEs, the majority population. Number of studies discussing the origins of these compact galaxies have reached to the conclusions that are contradictory to each other \citep{Trujillo07,Paudel14,Graham15,Stringer15}.

With advent of large scale surveys and high-resolution imaging instruments, an increasing number of cEs have been reported \citep{Chilingarian09,Chilingarian15}. They are found in a range of environments: from the densest galaxy cluster centre, to the field \citep{Mieske05,Huxor11,Huxor13,Paudel14,Price09}. However, an overwhelm majority of known cEs are located in dense environments, particularly around massive galaxies. Therefore, several studies propose an origin via tidal stripping, where strong tidal forces from a nearby massive host galaxy strips off the entire outer-disk component during the interaction and the remnant inner bulge  becomes a naked compact galaxy  \citep{Bekki01,Choi02}.  \cite{Huxor11} discovered two cEs, each near a massive host and with clear tidal debris,  confirming this picture. Other hand, cEs are also found in isolation or less dense environments  \citep{Huxor13,Paudel14,Chilingarian15} and in the absence of such strong tidal force, it is therefore naively expected that not all cEs might have formed via tidal stripping. \citet{Chilingarian15} has, however,  suggested that all such isolated cEs can be explained as runaway systems, ejected from a galaxy group or cluster by three-body encounters.

Here, we present a new compact galaxy, SDSS J1229+0001, which, from its  morphology, seems to be an early-type galaxy,  visually selected to create an early-type galaxy sample from the Sloan Digital Sky Survey (SDSS) colour images. The presence of strong emission in H$\alpha$ in optical spectrum, however, indicates a burst of ongoing star-formation similar to a typical Blue Compact Dwarf (BCD) \citep{Hopkins02}.

\section{Data analysis}
We perform a detailed analysis of imaging and spectroscopic data to characterise the stellar population and structural properties of this  galaxy. The multi-wavelength study covering the range from optical to far-infrared data, and based on various archival images, allows us to make an accurate estimation of stellar mass, dust mass and internal extinction. We obtained long-slit spectroscopic observation using ESO-VLT that supplements these data. Comparing the derived stellar population and structural properties with those of a well-studied sample of galaxies, we try to explore its possible evolution and origin which may provide some important clue in understanding of the formation and evolution of compact galaxies in the nearby Universe.

\subsection{Environment and Identification}\label{env}
\begin{figure}
\includegraphics[width=8.5cm]{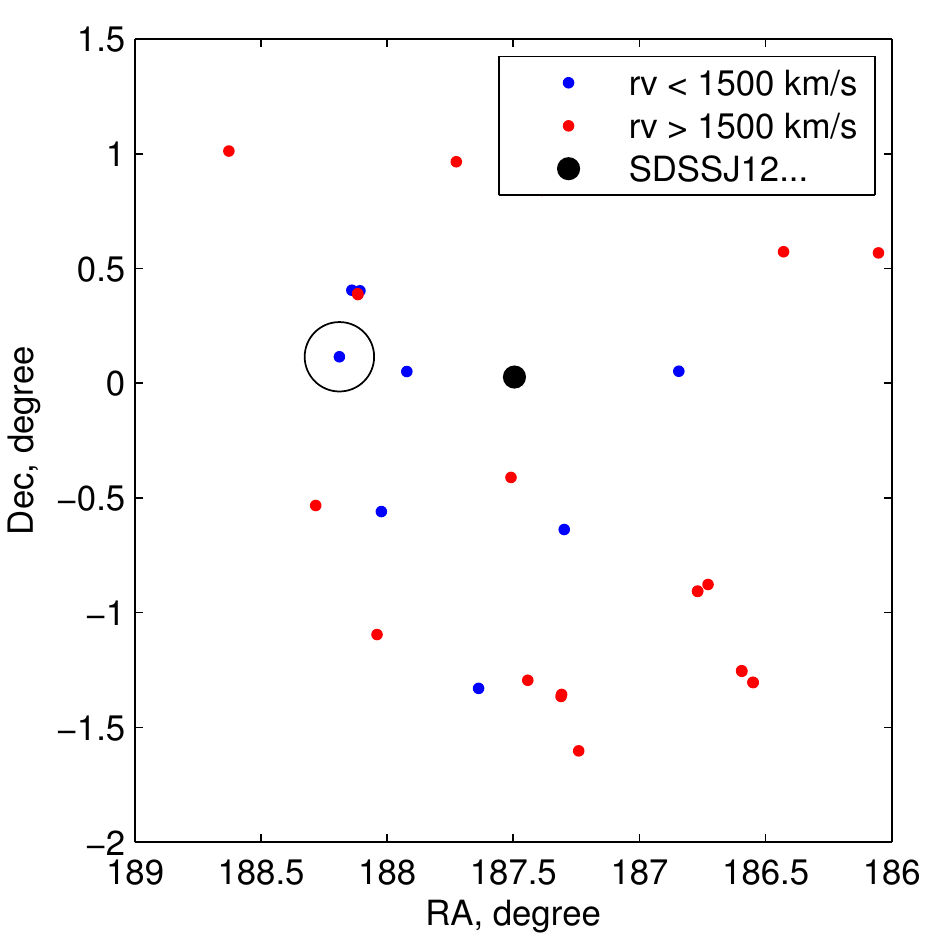}
\caption{Distribution of the line of sight radial velocities over a range of 3000 km/s of galaxies around SDSS J1229+0001. The red and blue dots represents galaxies which radial velocity larger and smaller than 1500 km~s$^{-1}$, respectively. A large solid black dot represents SDSS J1229+0001 and the black circle is the position of NGC 4517, where  the two have line of sight radial velocities of 2413  km~s$^{-1}$ and 1128 km~s$^{-1}$, resepctively. }
\label{alsky}
\end{figure}

We first conducted a systematic search of low-mass early-type galaxies in the field environment of the local volume (z $<$ 0.01) to catalogue and study the physical and stellar population parameters of those galaxies located in the isolation. For this purpose we derived the distance to the nearest massive galaxies in terms of sky projected separation and relative velocities \citep[see][]{Paudel14} and selected the early-type morphology by visual inspection of the SDSS DR7 colour images \citep{Abazajian09}.

SDSS J122958.84+000138.0 (hereafter SDSS J1229+0001) is located in a relatively isolated environment south of the Virgo cluster, at an angular separation of $\sim$8 degree from M49. The nearest bright galaxy is NGC 4517 is at $\sim$45' (440 kpc) east although the relative radial velocity between the two is 1285 km~s$^{-1}$. But the closest galaxy in term of sky-projected distance and relative radial velocity is SDSS J123002.09-002438.0; located at $\sim$190 kpc south and with a relative radial velocity\footnote{Redshift measured by 2DF survey \citep{Colless01} which report the line of sight radial velocities of SDSS J1229+0001 and it's companion 2408.1 and 2310.0 km/s, respectively.} 98 km~s$^{-1}$ but have a luminosity of M$_{r} = -$14.71 mag. 

In Figure \ref{alsky}, we show distribution of galaxy around SDSS J1229+0001 within a radius of 100' and redshift range of z = 0.001 to 0.01.  The galaxy sample is obtained from a NED query. The sample is divided into two redshift bins, i.e z $<$ 0.005 and z $>$ 0.005, represented by blue and red points respectively. SDSS J1229+0001 is represented by large solid dot and a big circle represents NGC 4517, having line of sight radial velocities of  2413  km~s$^{-1}$ and 1128 km~s$^{-1}$ resepctively. According to a catalogue of nearby groups \citep{Makarov11}, SDSS J1229+0001 is not assigned to be a member of any nearby group. Interestingly, NGC 4517 is also considered an isolated disk galaxy in \cite{Doyle05} and galaxy population in this region of sky seems to be relatively sparse. Within this the sky coverage shown in the Figure \ref{alsky}, no rich galaxy group has been found.

\begin{figure}
\includegraphics[width=7.5cm]{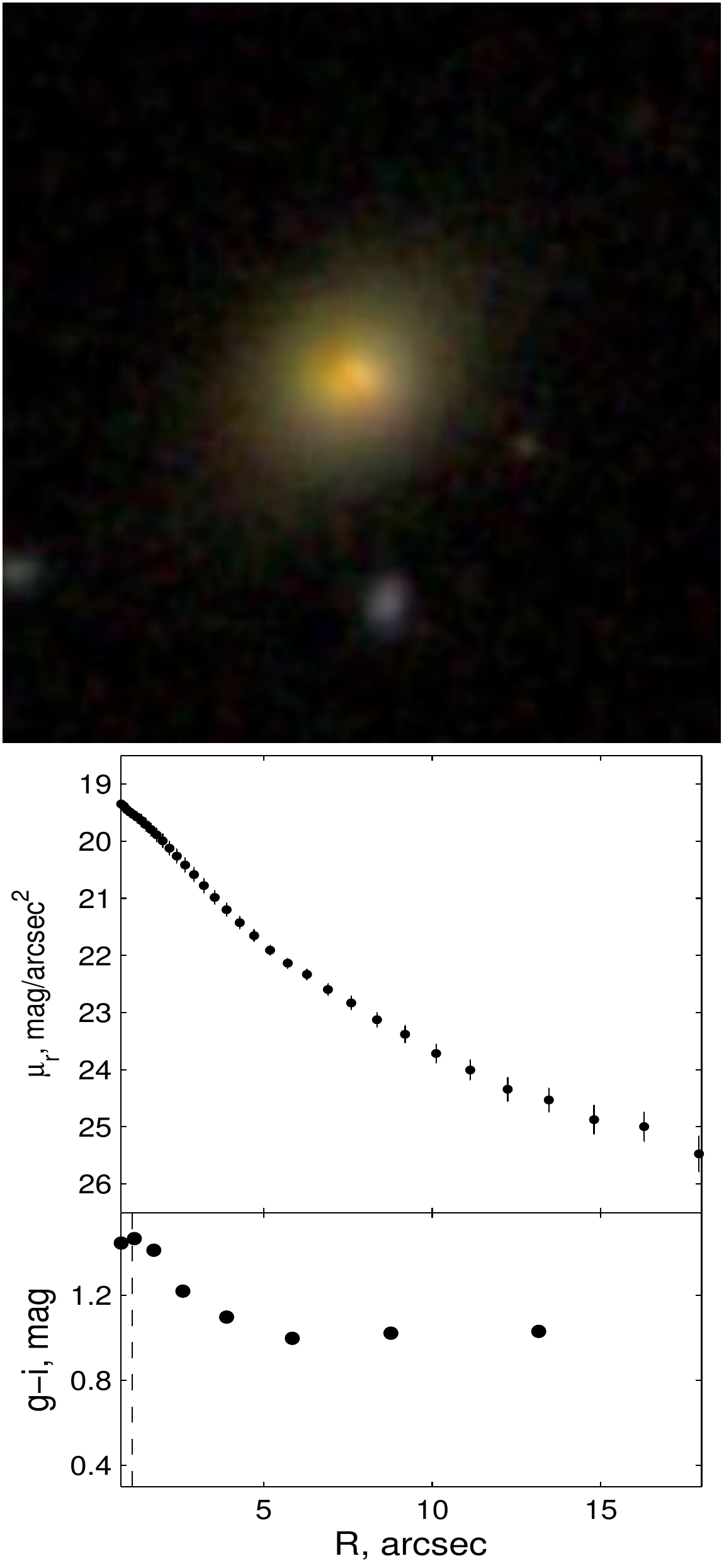}
\caption{Optical view of the SDSS J1229+0001 with a field of view of 1' $\times$ 1'.\newline
 Top: We show colour-image cutout from the SDSS\newline
Bottom: We show major axis profile measured from {\sc IRAF}/$ellipse$.  In second panel we show $g-i$ colour gradient (without corrections for either internal and external extinction). Where the vertical dash line represent the $r-$band PSF.  }
\label{prof}
\end{figure}

\subsection{Imaging}\label{imaging}
At first, we used the SDSS image to perform image analysis. We particularly made use of $r-$band image, which has a higher signal to noise ratio than the other bands. The archival images of SDSS-III has better sky-background subtraction than the previous data releases. We, further, subtracted the sky-background using the procedure in \cite{Paudel14} where we selected the sky-background counts from randomly selected sky-regions around the galaxy to derive the median background. 

The IRAF task $ellipse$ was used to extract the galaxy's major-axis light profile. We used similar procedure as in \cite{Paudel14} to prepare input image and to run ellipse task. In Figure \ref{prof}, we show $g-r-i$ combined colour image cut-out in top panel which is we obtained from the SDSS sky-server. At first glance, it gives a visual impression of S0  galaxy with a prominent dust lane at the centre, perpendicular to the major axis of the galaxy. We show the major axis light profile and $g-i$ colour profile in bottom panel of Figure \ref{prof}. The colour gradient was derived from the azimuthally-averaged light profiles in the $g$ and $i$-band. Since a difference in PSF  can produce an artificial colour gradient at the centre of galaxies, we matched the PSFs of these images by degrading the better image. We find that PSFs in $g$ and $i$-band  are 1.04$\arcsec$ and 0.92$\arcsec$, respectively.

\begin{figure}
\includegraphics[width=8.5cm]{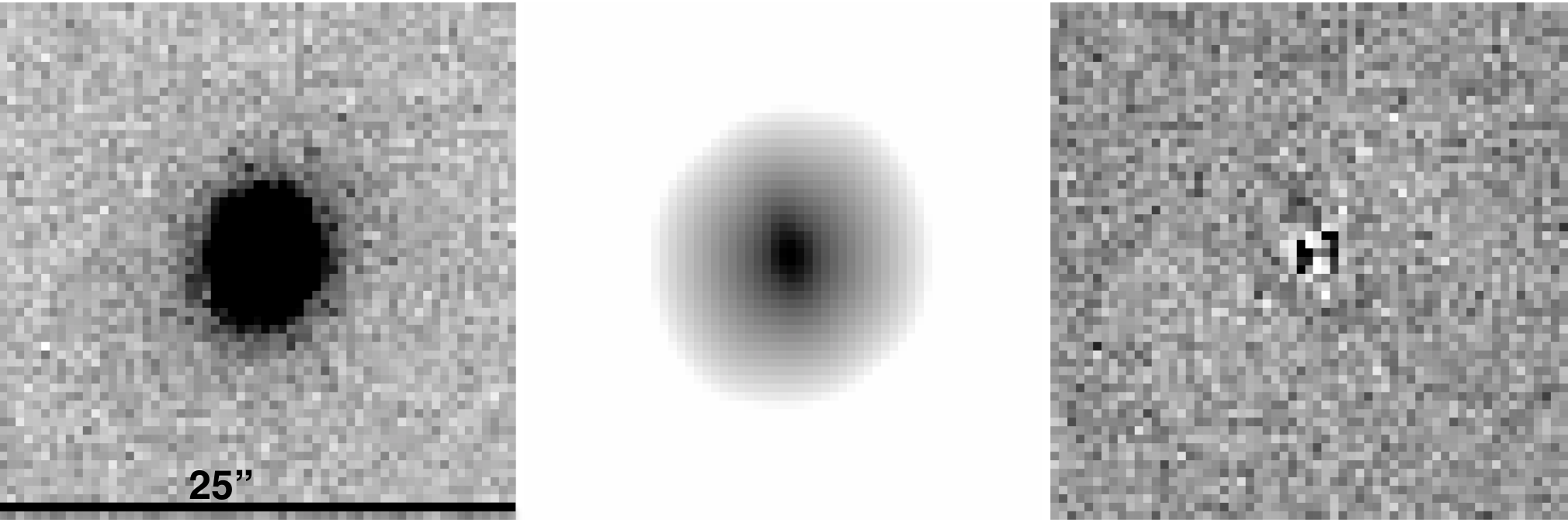}
\includegraphics[width=8.5cm]{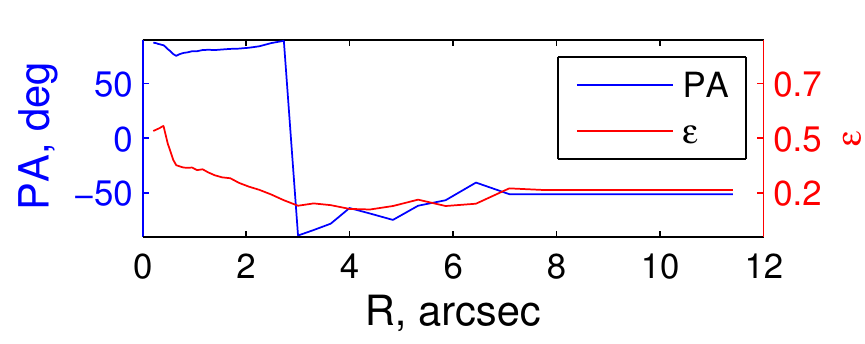}
\caption{Modelling of galaxy light distribution with {\sc galfit} . We show the H-band image, the model and the residual in left, middle and right panel, respectively. We show variation of ellipticity (red) and PA (blue) along the major axis in bottom panel.  }
\label{galfit}
\end{figure}

Since the central part of the galaxy heavily obscured by dust, to model the galaxy light distribution and derive structural parameter we used UKIDSS\footnote{http://www.ukidss.org} H-band image expecting that the H-band image is less affected by dust extinction. We retrieved stacked H-band image from WSA - WFCAM Science Archive\footnote{http://wsa.roe.ac.uk} which has spatial sampling of 0.4" and the spatial resolution is comparable to the SDSS with a seeing of 0.84". We used {\sc galfit} to perform two dimensional modeling. The required PSF image is also provided during {\sc galfit} run. We find that the two dimensional light distribution is better modeled with two component S\'ersic function. We show result of {\sc galfit} run in top panel of Figure \ref{galfit}. Where we show, form left to right, the H-band image, the model and the model subtracted residual, respectively. We also run {\sc iraf} $ellipse$ task in H-band image to derive variation of ellipse parameter along the major axis that we show in lower panel of Figure \ref{galfit}.  The average ellipticity of the galaxy is 0.21.

Non parametric approach has been used to calculate photometric parameters, i.e half-light radius and total luminosity. Where a two Petrosian radius aperture has been used to summed up the total flux. We follow a similar procedure as described in \cite{Janz08,Paudel14} where we used major axis Petrosian radius. The major axis half-light radius is circularized by multiplying a factor of (1-$\epsilon$)$^{0.5}$ where $\epsilon$ is an average ellipticity of the galaxy. The mean surface brightness within the half-light radius is calculated by using the equation $<\mu> = m+2.5log[2\pi R_{h}^{2}]$, where R$_{h}$ is half-light radius in arcsec. The results are presented in Table \ref{par}, where the magnitudes are corrected for both Galactic and internal extinction. The value of Galactic extinction is obtained from NED and the internal extinction due to dust is derived from the SED fitting (see below).  

In this work, adopted a cosmology with H$_{0}$ = 71 and $\Omega_{m}$ = 0.3. Using the Hubble flow we obtained a distance to the galaxy is 34.0 Mpc and spatial scale 0.163 kpc arcsec$^{-1}$.

Fortunately, we find that substantial multi-wavelength data are available in various archives, allowing us to create a Spectral Energy Distribution (SED)  from the optical to the far infra-red (FIR). The SDSS provides optical --$u, g, r, i$ and $z$-bands images. We obtained corrected UKIDSS J, H and K band magnitudes from RCSED\footnote{ http://rcsed.sai.msu.ru}. We use the IRSA\footnote{http://irsa.ipac.caltech.edu} archive to obtain NIR images (Spitzer observations) and measure the total flux using aperture photometry. The CDS\footnote{http://cdsarc.u-strasbg.fr} cataloge server is queried to acquire the MIR and FIR fluxes.  Where 24$\mu$, 12$\mu$ and 22$\mu$ fluxes were measured in the Spitzer MIPS24, WISE3\footnote{http://wise.ssl.berkeley.edu} and WISE4 imaging bands, respectively. The FIR data points are taken from the Infrared Astronomical Satellite (IRAS) and AKARI\footnote{http://www.ir.isas.jaxa.jp} all sky surveys. The measured fluxes in respective wave-bands in Janskys are given in Table \ref{multi}. 

\begin{table}
\caption{Multi-wavelength data}
\begin{tabular}{lcr}
\hline
Filters & Lambda\_eff & Flux \\
 & $\mu$ & jy \\
\hline
SDSS $u'$ &  0.35   &   2.4E-4   $\pm$ 0.5E-5 \\
SDSS $g'$ &  0.46   &   9.42E-4  $\pm$ 0.5E-5 \\
SDSS $r'$ &  0.61   &   2.07E-3  $\pm$ 0.5E-5 \\
SDSS $i'$ &  0.74   &   3.36E-3  $\pm$ 0.5E-5 \\
SDSS $z'$ &  0.89   &   3.90E-3  $\pm$ 1E-5  \\
UKIDSS J &  1.25   &   8.44E-3  $\pm$ 1E-5  \\
UKIDSS H &  1.65   &   11.35E-3  $\pm$ 1E-5  \\
UKIDSS K &  2.20   &  9.68E-3  $\pm$ 1E-5  \\
IRAC1   &  3.56   &   4.83E-3  $\pm$ 4E-5  \\
IRAC2   &  4.51   &   3.54E-3  $\pm$ 3E-5  \\
IRAC3   &  5.76   &   13.36e-3 $\pm$ 1E-4  \\
IRAC4   &  8.00   &   39.18e-3 $\pm$ 1E-4  \\
WISE3   &  12.0   &   24.20E-3 $\pm$ 1E-3  \\
WISE4   &  22.0   &   62.11E-3 $\pm$ 1E-3  \\
MIPS24  &  24.0	  &    0.06    $\pm$ 1E-3  \\
IRS60   &  60     &   0.81     $\pm$ 2E-3  \\
IRS100  &  100    &   1.08     $\pm$ 5E-2  \\
AKARI140  &  140    &   0.38     $\pm$ 1E-1  \\
AKARI160  &  160    &   0.20     $\pm$ 1E-1  \\
\hline
 \end{tabular}
 \label{multi}
 \end{table}

The publicly available code MAGPHYS \citep{Cunha08} was used to analyse the observed SED. The main principle behind it, in brief, is to check the energy balance that are emitted from several different components, i.e. young and old stellar emission, absorption due to dust and re-emission in FIR. First a model SED is created using \cite{Bruzual03} library of templates computing integrated stellar light and superposing an attenuation curve due to dust for a varying set of model parameters such as SFR, age and metallicity.  The dust attenuation curve is derived from a simple analytic model of \cite{Charlot00}. Finally, applying a $\chi^{2}$-minimisation scheme, MAGPHYS tries to find the best fit model template for a particular set of physical parameters. This code does not consider a potential AGN contribution and only star formation processes are involved in building the SEDs. However, we investigate any possibility of AGN presence by looking its position in BPT diagram \citep{Baldwin81}. We use the emission line flux measured from central spectrum to construct the BPT diagram, and find that SDSS J1229+0001 is unambiguously located in the region of star-forming galaxies. 

\begin{figure*}
\includegraphics[width=18cm]{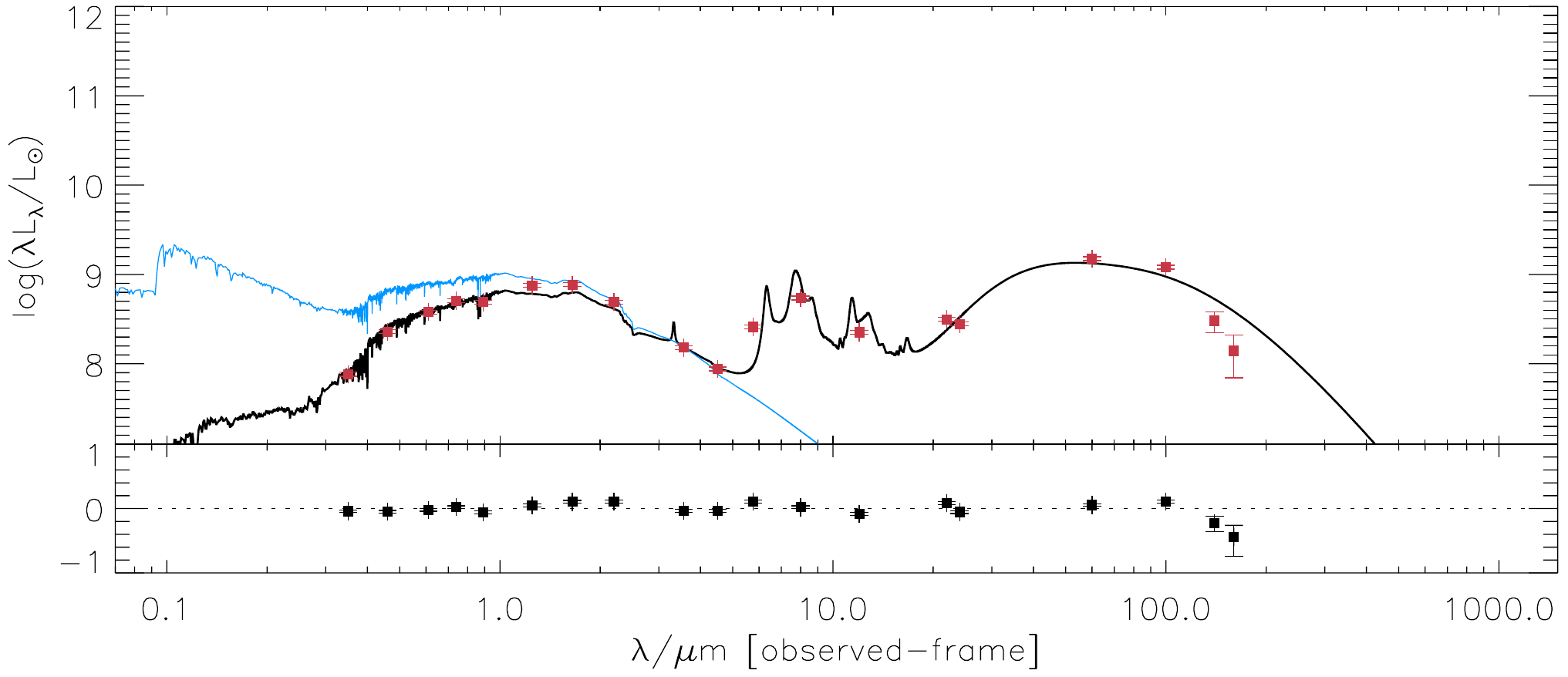}
\caption{Fitted SED, output from MAGPHYS. The red dots are observed data points of SDSS J1229+0001. The black and blue lines are the best fit SEDs for the non-attenuated and dust-attenuated model, respectively. }
\label{sed}
\end{figure*}

A comparison between the observed SED and the best fit template is shown in Figure \ref{sed}, where the red dots represent the observe photometric data points and the black and blue lines are the best fit SEDs for the non-attenuated and dust-attenuated model, respectively. In general, the model SED seems to be able to well characterise the observed data points. A slight deviation, however, is seen in the FIR region that may indicate the necessity of longer wavelength data points, unfortunately not available, to better constrain the physical parameters of the model. In particular, longer wavelength data are crucial in making an accurate estimate of dust temperature. 

The resulting physical parameters: total stellar mass, dust mass, extinction and metallicity are reported in last panel of Table \ref{par}. We derive the SDSS $r$-band extinction, A$_{r}$, from the flux ratio between the dust-attenuated and non-attenuated model i.e. the black and blue line in Figure \ref{sed}, respectively.

\subsection{Spectroscopy}\label{specdata}

\begin{figure}
\includegraphics[width=8cm]{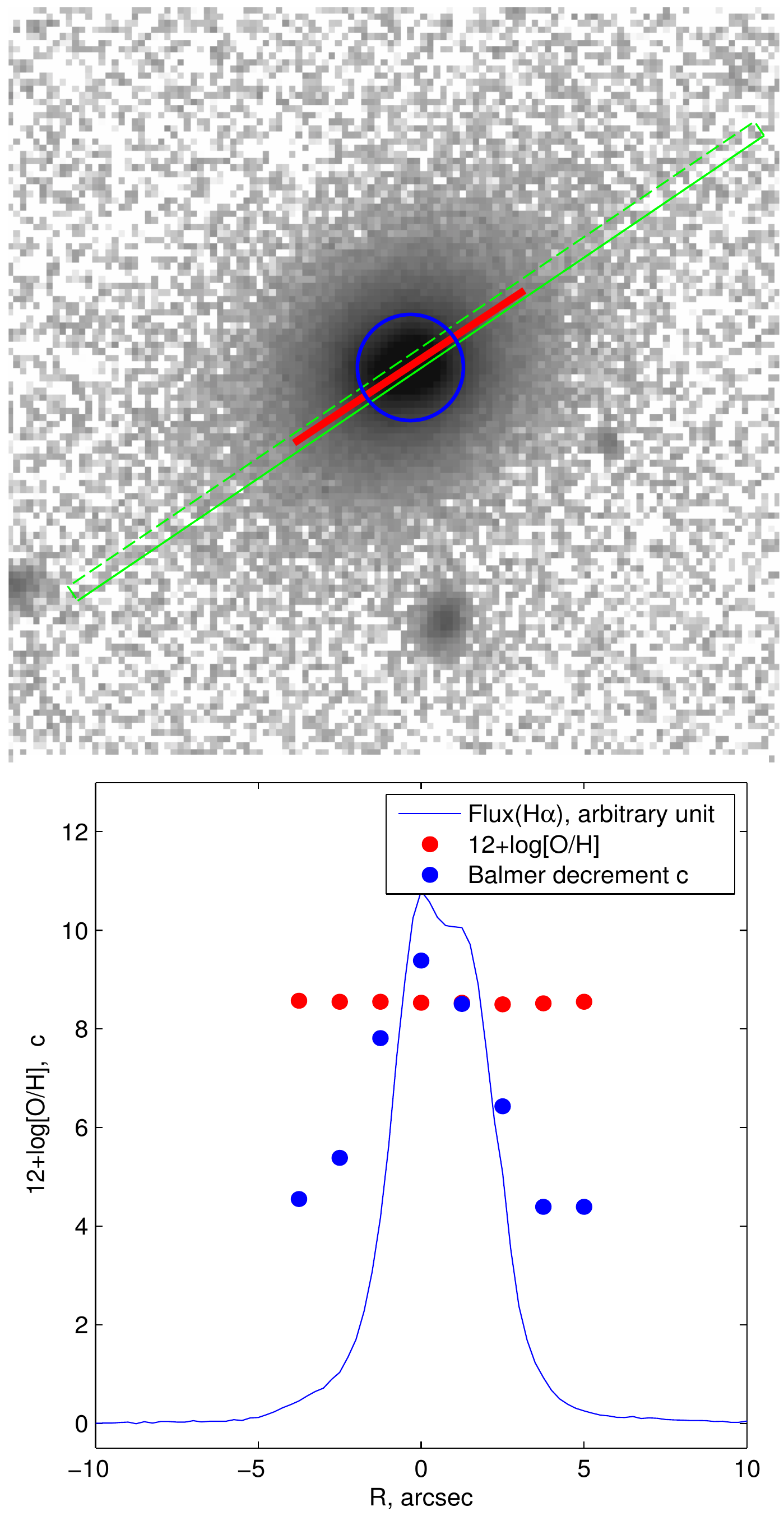}
\caption{Top: SDSS r-band image where we overlay a one arcsecond slit in green colour. The red solid line represents the extension the extension of H$\alpha$ emission and the radius blue circle is half-radius of galaxy. The field of view is similar to Figure 2.
\newline
Bottom: H$\alpha$, gas-phase metallicity and Balmer decrement (c) profile along the major axis. }
\label{slit}
\end{figure}

We obtained long-slit low-resolution spectroscopic data using VLT FORS2 instrument as a part of ongoing campaign to study low-mass early-type galaxies in different environment. The details of long-slit observation and data reduction were presented in \cite{Paudel14}. The total exposure time for this galaxy was 1000 sec. Following the similar procedure of VLT FORS2 data reduction as described in \cite{Paudel10} we achieve 0.1 mag accuracy around 5000 \AA{} in calibrating the observed flux that we measured from standard deviation of sensitivity function. The orientation and position  of the slit over the galaxy is shown in Figure \ref{slit}, top panel where the slit is aligned along the major-axis of the galaxy.

\begin{figure}
\includegraphics[width=8.5cm]{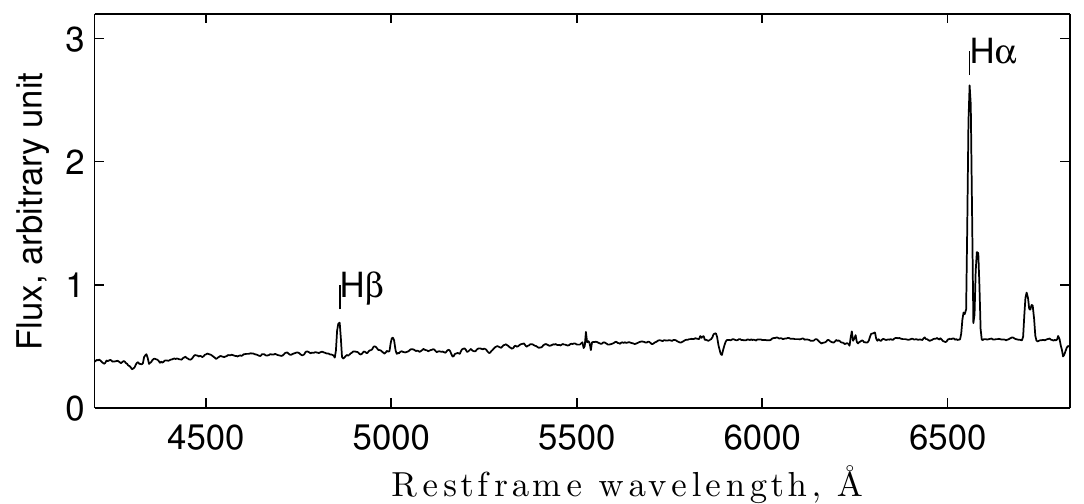}
\caption{Flux-calibrated VLT2 FORS2 spectrum of SDSS J1229+0001.}
\label{spectrum}
\end{figure}

The one-dimensional spectrum is extracted by summing the flux in a two-dimensional frame along the spatial direction. To extract a galaxy wide  spectrum we use an aperture of $\pm$7$\arcsec$ (solid red line in Figure \ref{slit} top panel). In Figure \ref{spectrum}, we show the flux-calibrated one-dimensional spectrum, where the Balmer emission lines are prominent.

We measure the emission line fluxes using the IRAF task $splot$ fitting a gaussian profile. The ratio between H$\alpha$ and H$\beta$ flux,  so called Balmer decrement c, is  7.8. Using a theoretical value of the c0 = 2.86 for a electron temperature 10$^{4}$ K and an electron density n$_{e}$ = 10$^{2}$ cm$^{-3}$ \citep{Osterbrock89} and the \cite{Calzetti00} extinction law, we derived the extinction coefficient E(B-V) and corresponding r-band extinction A$_{r}$.

Oxygen abundances, 12+log(O/H), were estimated with the  two  methods described  by \cite{Marino13}, i.e the so-called N2 and O3N2 methods. The N2 method only considers the line ratio between H$\alpha$ and [NII]  while the O3N2 method uses a combination of the line ratios H$\alpha$/[NII] and [OIII]/H$\beta$. A typical systematic error in these calibrations is 0.2 dex \citep{Denicolo02,Marino13}.

\section{Results}

\begin{figure}
\includegraphics[width=8cm]{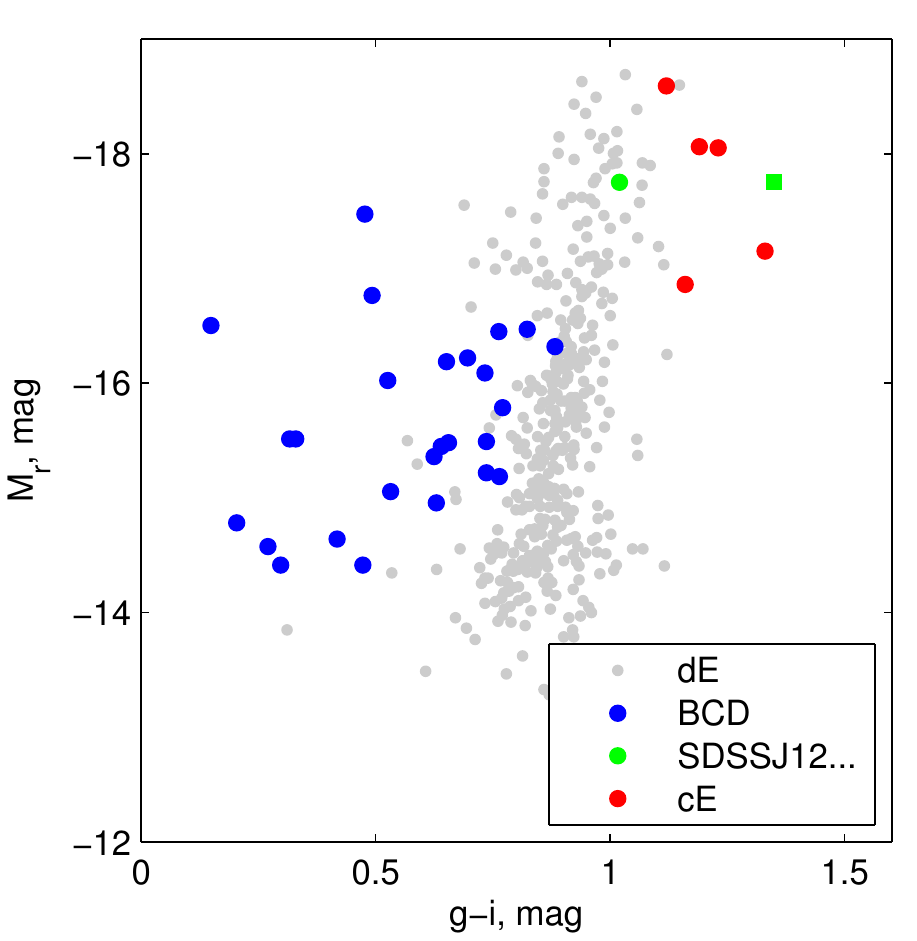}
\caption{Colour-magnitude relation of dwarf galaxies. BCDs, dEs  and cEs are in blue, grey and red, respectively. The green symbols represent  SDSS J1229+0001, where the square and circle are for before and after extinction correction, respectively.  }
\label{cmr}
\end{figure}

\begin{figure}
\includegraphics[width=8.5cm]{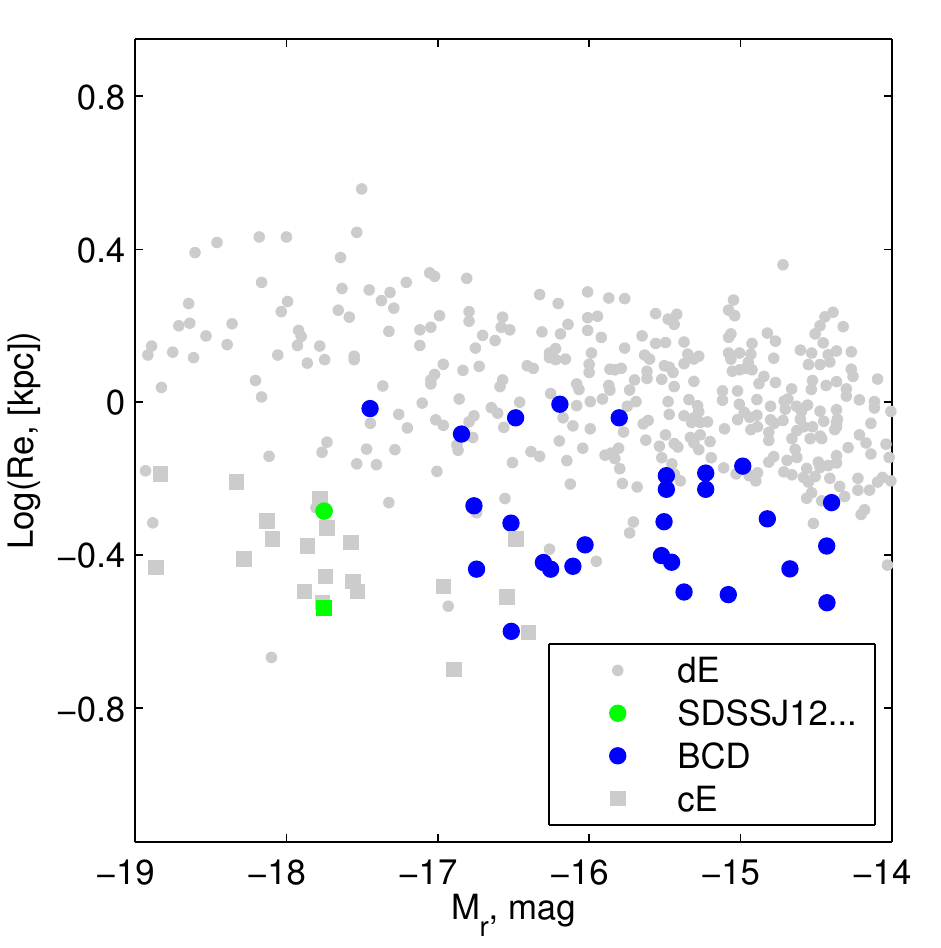}
\caption{Magnitude-size relation. The green circle and square represent the half-light radius of SDSS J1229+0001 derived from optical and H-band images, respectively. The other of symbols are similar to Figure \ref{cmr}.}
\label{msr}
\end{figure}

\begin{table*}
\label{par}
\caption{Global properties:
Physical parameter of SDSS J1229+0001. The parameters are grouped according to the results of different analyses.  Column 2 -- 4 are the results of the SDSS r-band image analysis. The magnitudes are corrected for galactic and internal extinction. R$_{h}$ is half light radius derived from Petorsian method. Emission line metallicity, 12+log(O/H), is listed in column 6. The output of the best-fit SED using MAGPHYS are presented in columns 7 -- 11. \newline
$^{*}$ Before extinction correction.
}
\begin{tabular}{lcccc|crr|rrrr}
\hline
 &  M$_{r}$ & $g-i$ & Re   &   z    & 12+log(O/H) & E(B-V)&M$_{*}$ & M$_{dust}$ & A$_{r}$ & log[Z/Z $_{\sun}$]\\
 \hline
 &  mag & mag & kpc &   & dex & mag & M$_{\sun}$ & M$_{\sun}$ & mag & dex\\
 \hline
SDSS J122958.84+000138.0 &  -17.75 &  1.02 (1.35$^{*}$) & 0.52  &  0.008   &8.5 & 0.85 &1.8E9 & 5.1E5 & 0.71 & 0.008\\
\hline
 \end{tabular}
  \end{table*}

Although a typical early-type galaxy as seen in the SDSS colour image, the optical spectroscopy  reveals strong Balmer emission (Fig. \ref{spectrum}). The star-formation rate derived from H$\alpha$ emission is $\sim$0.25 M$_{\sun}$/yr and, not surprisingly, as a sign of star-burst activity; the dust lane at centre is clearly visible.  Below we present the results from analysis of structural properties and multi-wavelength photometry.

 \subsection{Image analysis}
We find a significant difference in half-light radius measured in optical and H-band imaging. The measured geometric half-light light radius of the galaxy from the SDSS r-band is 0.52 kpc whereas that we measured from H-band imaging is significantly less i.e 0.29 kpc. The SDSS z-band the half-light  radius is 0.49 kpc. This trend can be well explained by increasing extinction of dust at the centre of the galaxy toward shorter wavelength which significantly reduces the central light concentration.

\begin{table}
\caption{Derived structural parameter from H-band image. All parameters, other than top three, are the results of {\sc galfit} run. m$_H$ is corrected H-band magnitude listed in RCSED and R$_{h}$ is half light radius measured from Ptrosian method, see text. $<\mu>_{H}$ is an average surface brightness within R$_{h}$.}
\begin{tabular}{ccc}
\hline
Parameter & value & unit\\
\hline
m$_H$ & 14.39 & mag\\
R$_{h}$ & 0.29 & kpc\\
$<\mu_H>$ & 17.53 & mag arcsec$^{-2}$\\
R$_{e,in}$ & 0.19 & kpc\\
R$_{e,out}$ & 0.33 & kpc\\
n$_{in}$ & 1.4 & \\
n$_{out}$ & 1.2 & \\
$\epsilon_{in}$ &0.62 & \\
$\epsilon_{out}$ &0.08 & \\
PA$_{in}$ & 77 & degree \\
PA$_{out}$ & -61 & degree  \\
\hline
\end{tabular}
\label{galpar}
\end{table}

It was clear from variation of position of PA in inner and outer part that the morphology of the galaxy is not simple. This was further supported by our two-dimensional modelling of light distribution in H-band image using {\sc galfit}. We find that the two component S\'ersic function provides better fits than simple S\'ersic function. The results are listed in Table \ref{galpar}. The best fit inner and outer component S\'ersic indices (n) are 1.4 and 1.2 and effective radii (Re) 0.19~kpc and 0.33~kpc respectively. The inner and outer component has Position Angle 77$^{\circ}$ and -61$^{\circ}$, respectively. We plot ellipcity and position angle profiles along the major axis in the last panel. We find strong change in PA around 2 to 3 arcsec which flips from $\sim$80$^{\circ}$ to $\sim$-80$^{\circ}$, Figure \ref{galfit} bottom panel.

An overall $g-i$ colour index of the galaxy is 1.02 and 1.35 mag, after and before applying the correction for extinction due to dust reddening, respectively. The relation between colour magnitude for various types of low mass galaxies is shown in Figure \ref{cmr}. For this we use tabulated values of $r$-band absolute magnitude and $g-i$ colour index from \cite{Meyer13} for BCDs and dEs. For the cEs, we use the photometric parameters from \cite{Chilingarian08,Chilingarian10}.  Two green symbols, solid circle and square, represent the extinction corrected and uncorrected $g-i$ colour index of SDSS J1229+0001, respectively.  The difference between star-forming galaxies, BCDs, and others (cEs and dEs) is distinctively clear as previously noted  by \cite{Meyer13}, but interestingly SDSS J1229+0001 is consistent with dEs colour being slightly bluer than mean cE colour. Nevertheless, the uncorrected $g-i$ colour of SDSS J1229+0001 is the reddest point in this diagram.

We show the observed $g-i$ colour gradient (uncorrected for extinction) along the major axis in Figure \ref{prof}, where we can see that steep rise of $g-i$ colour index at the centre with a maximum value 1.4 mag. It becomes nearly constant beyond the 3$\arcsec$ and have a value similar to that we obtained from the overall colour index after the extinction correction that is obtained from SED fitting. 

In Figure \ref{msr}, we show the relation between sizes and luminosities for the low mass early-type galaxies, using the measurements from \citet{Janz08} for  dEs, Es and S0, and Chilingarian et al. for cEs. As previously identified in many studies \citep{Janz08,Chilingarian09,Huxor13,Paudel14}, the distinction between cEs and dEs in this diagram is clearly visible. SDSS J1229+0001 lies in the region of cE being relatively compact compare to similar luminosity dEs or dS0. Note, however, there is also statistical difference in sizes of BCDs and dEs as the latter being more extended system then former, though the scatter in both cases is fairly large.

\subsection{Multi-wavelength analysis}
The results from the analysis of multi wavelength data, i.e a SED from optical to FIR, are presented in Table \ref{par}, fourth panel. The derived total stellar and dust masses are 1.8 $\times$ 10$^{9}$  and 5.1 $\times$ 10$^{5}$ M$_{\sun}$, respectively which gives a value of dust-to-stellar mass ratio, log(M$_{dust}$/M$_{*}$) = $-$3.5. The metallicity derived from SED fitting agrees well with that from emission lines, being nearly solar at log(Z/Z$_{\sun}$) = 0.008. However, the internal extinction derived from flux ratio between the Balmer lines is significantly larger than than that we have obtained from SED fitting. Comparing the attenuated and non-attenuated best fit SED, we calculate extinction $A_{\lambda} = 2.5log(\frac{F_{\lambda}^{non-attenuated}}{F_{\lambda}^{attenuated}})$. We obtain A$_{r}$ = 0.7 mag, at wavelength 7500 \AA, and the extinction becomes as high as A$_{uv}$ = 5.9 mag at Ultra Violet (UV, $\lambda$ = 1000 \AA) region. We expect this difference in extinction which are derived from different methods because the Balmer decrement only represent the central star-forming region where the presence of dust is pronounce and in the SED fitting we consider overall galaxy photometry.

We derived total star-formation rate from several different diagnostics and respective empirical calibrations. These diagnostic are listed in Table \ref{sfr} where the fluxes are presented in Jansky unit. H$\alpha$ emission line flux is corrected for the internal extinction, but no aperture correction has been made. We use empirical calibrations from \cite{Kennicutt98,Murphy11,Rieke09} for the H$\alpha$, radio and MIPS star-formation diagnostics, respectively. We find that the SFR derived from these three diagnostics are fairly consistent with each other but  SED fitting gives a lower value, i.e nearly four time smaller than that from H$\alpha$ or 1.4GHZ fluxes. Due to uncertainties in the model, particularly the star-formation history, it is likely that SFR derived from SED fitting without UV photometry is highly uncertain. 

 \begin{table}
 \label{sfr}
 \caption{Star-formation rates}
\begin{tabular}{llr}
\hline
Band & Flux & SFR \\
\hline
 & Jy & M$_{\sun}$/yr\\
\hline
H$\alpha$  & 0.34$^{*}$ $\pm$ 0.005 & 0.26\\
MIPS24$\mu$ & 0.06 $\pm$0.001 & 0.19 \\
1.4GHz & (2.9 $\pm$ 0.13)$\times$10$^{-3}$  & 0.25\\
SED fitting & - & 0.06 \\
\hline
 \end{tabular}
 {\\
 Star-formation rates calculated from different methods. \\
 $^{*}$The H$\alpha$ flux is obtained by summing up the long-slit spectrum and finally corrected for extinction A$\alpha$ = 2.3 mag.}
 \end{table}

\subsection{Spectroscopy}
 
We show H$\alpha$ flux profile along the slit in Figure \ref{slit}, blue line at lower panel. We see nearly two peaks gaussian profile that may hint at some mis-alignment of the elongation of star-forming region at the centre with PA of the galaxy. This is much clearer in the colour image, where the dust lane is almost orthogonal to this PA  and probably why the PA profile shows an abrupt change at the centre, see Figure \ref{prof}. The extension of H$\alpha$ emission along the major axis is greater than $\pm$6$\arcsec$ from the centre.

The total H$\alpha$ flux measured from integrated spectrum is 2.66E-14 erg/cm2/s\footnote{Without correcting the extinction and aperture effects}. The ratio between H$\alpha$ and H$\beta$ flux, Balmer decrement, c = 7.8. Which gives E(B-V)= 1.97 $\times$ log(c/c0) = 0.85 and corresponding $r$-band extinction A$_{r}$ = 3.44 mag. However note that, we do not use this value of A$_{r}$ to correct any photometric measurement other than the H$\alpha$ flux to derive star-formation rate, see below.

We obtained 12+log(O/H) = 8.4(8.5)  dex from the N2(O3N2).  We show metallicity profile along the slit in Figure \ref{slit}, lower panel red-solid dots. We find a nearly constant gas-phase metallicity along the major axis up to $\sim$ 2 R$_{e}$

\section{Discussion and Conclusion}
We present a new compact and red looking star-forming galaxy, SDSS J1229+000, which has a half-light radius of 520 pc and relatively steep light profile with S\'ersic index n = 3. Although, SDSS J1229+0001 has a $g-i$ colour index of 1.02, similar to a typical old ``red and dead" early-type galaxy of a similar mass, a strong ongoing star-formation is also observed at the centre, with a SFR = $\sim$0.25 M$_{\sun}$/yr. Nevertheless, an analysis of SED from optical to FIR reveals that a substantial amount of dust is present, i.e. M$_{dust}$ =  4.9 $\times$ 10$^{5}$ M$_{\sun}$ with dust-to-stellar mass ratio of log(M$_{dust}$/M$_{*}$) = $-$3.5.

\subsection{Mixed/Intermediate morphology?}
Although low mass galaxies are the most common in the universe, their classification scheme is somewhat arbitrary. Particular morphological traits, mainly in the optical bands, are used to define a particular class of galaxies. For example, strongly concentrated star-formation is mostly considered for  Blue Compact Dwarf galaxy (BCD) and irregular morphology with low level of star-formation are characteristics of dwarf Irregular galaxies (dIrrs). They are both gas rich and found in relatively low dense environments, such as in the outskirts of galaxy-cluster and in the field. BCDs are, by definition, relatively high surface galaxies compared to  dIrrs.  \cite{Meyer13} showed that a number of BCDs (but not all) are structurally similar to compact early-type dwarf galaxies (dEs) and that dIrrs are similar to diffuse dEs.
 
Among the non star-forming families there exist several types. dEs, dwarf lenticulars (dS0s) and dwarf spheroidals (dSphs) are relatively extended galaxies with low-surface brightnesses. The distinction between them is not straightforward. However, the primary motivation of introducing the dS0 class was the observation of multi-component light profiles in some low mass early-type galaxies in the Virgo cluster \citep{Sandage84} --   dwarf analogs of the massive S0 galaxies which posses distinct bulge and disk components. Furthermore, \cite{Lisker07} detailed analysis of substructural properties of dEs indeed supports the existence of disk feature in a considerable fraction of dEs. \cite{Janz14} concluded that the common bulge-plus-disk picture of bright S0s is not applicable, explaining the observed inner and outer components of early-type dwarfs.  Moreover,  \cite{Ryden99} earlier  noticed no distinction in the structural parameters of these two --dEs and dS0-- morphological classes. However, the average dS0 has a brighter surface brightness than the average dE,  reflecting the fact that dS0s mostly occur at brighter magnitudes of early-type dwarfs. 

At the massive scale, S0 galaxies not only differ from Es in structural parameters but also in stellar population and dust content. More S0 galaxies host ongoing centrally concentrated star formation  than Es and also a significantly larger dust mass fraction \citep{Amblard14}. There are low-mass S0s which have been explicitly classified as S0 at the same mass as dS0s, but which have significantly higher surface brightness, e.g VCC 1833 or VCC 0140. The dust to stellar mass ratio of log(M$_{dust}$/M$_{*}$) = $-$3.5 is typical of early-type galaxies at a  stellar mass of $\sim$ 10$^{9}$ M$_{\sun}$ \citep{Alighieri13}. SDSS J1229+0001 also exhibits an unusually high star formation rate and dust mass that compared a typical S0 galaxy. In addition to this, we find that observed light distribution is better described by two component S\'ersic function. Overall size, half-light radius, measured in H-band is quite small, i.e 290 pc. Given that the H-band is more sensitive to the light of old stars than the optical is, this indicates that the galaxy will likely end up in the compact regime once star formation stops.

 Compact non-star forming galaxies, somewhat rare in observation, also exist in different names at different stellar masses such as compact ellipticals (cEs) and Ultra Compact Dwarf galaxies (UCDs) possibly form a continuum of a mass range from $\approx$10$^{6}$M$_{\sun}$ to $\approx$10$^{9}$M$_{\sun}$ \citep{Norris14}. Compact ellipticals (cEs) are significantly compact and high surface brightness galaxies compare to low mass S0s and dS0s. 

While not denying the possibility of it being a S0 galaxy, our results suggest that SDSS J1229+0001 may be the late stage of a forming compact early-type galaxy. It is worth noting that cEs and S0 morphologies are not mutually exclusive. M32, the type example of an cE, also possesses outer-disk and inner-bulge component which has been a defining character of S0 morphology. Recently, \cite{Paudel16} reported a compact early-type galaxy with an active nucleus which is structurally similar to a typical S0 galaxies.

\subsection{Relative rarity of the object }
In this section, we attempt to at least qualify how rare the compact, red-looking, star-forming galaxies are.  Both stellar population evolution and dynamical evolution of any compact young stellar system will lead to an expansion in size and dimming in brightness. The observation of compact and old non-star-forming objects means that once they were much more compact and brighter while they were young --  if they  formed in isolation. There exist the compact star-forming galaxies, so-called Blue Compact Dwarf galaxies (BCDs), some of them are as compact as cE, but they are less luminous than cEs. Note that all BCDs are not, in general, compact galaxies. In most cases, they are BCDs due to presence of blue compact star-forming region on the top of underlying extended relatively old component. The brightest BCD in the Virgo cluster, VCC 324,  has a total luminosity of  M$_{r}$ =  -17.46 mag and overall effective radius of 0.96 kpc. Selecting compact BCDs of sizes similar to cEs from \cite{Meyer13} sample we find a clear offset in magnitude distribution where cEs are on average a magnitude brighter than compact BCDs.

\begin{figure}
\includegraphics[width=8cm]{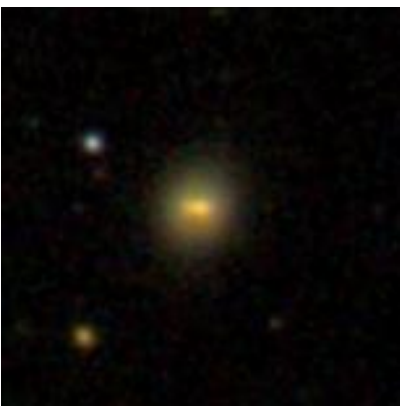}
\caption{The SDSS colour image of SDSS J112308.77+624845.6 with a field of view 1$\arcmin$ $\times$ 1$\arcmin$.}
\label{rcd2}
\end{figure}

To get a more statistically complete view of compact galaxies like SDSS J1229+0001 we visually analyse morphology and $g-i$ colour index of  a sample selected from SDSS. We select the galaxies sample from the SDSS catalogue within the redshift range z $<$ 0.02\footnote{We have chosen this redshift range because beyond 0.02 the morphological classification may not be reliable and also beyond this with a typical SDSS PSF of 1$\arcsec$ it is not sufficient to measure the R$_{e}$ of these compact galaxies.} which have effective radii (in the SDSS catalogue it is Petrosian half-light radius) $<$ 0.6 kpc\footnote{There are $\approx$500 galaxies and a significant majority of them are BCD type.}.

We find only one additional example, SDSS J112308.77+624845.6 a very similar morphology galaxy, see Figure \ref{rcd2}. Which have a Petrosian half light radius, measured by the SDSS, of 534 pc. The $g-i$ colour, before correcting the internal extinction, index is 1.3 mag and emission line metallicity, 12 +log(O/H), is 8.54. We find that the $g-i$ colour index larger than 1.3 mag is an extreme for similar mass galaxies and interestingly non of Virgo cluster galaxies, of both star-forming or non star-forming types, have that $g-i$ colour index.

\subsection{Star-Formation and Evolution}
The detection of star-formation activity in early-type galaxies is not new, nor is the presence of significant amount of neutral Hydrogen (HI). A dedicated study of HI distribution in Es by ATLAS-3D team detected HI in 40 percent of Es outside a cluster environment \citep{Serra12}.  Recent data release of  HI survey by ALPHA-APHA team reported that SDSS J1229+0001 has total HI mass of 4$\times$10$^{8}$ M$_{\sun}$ and corresponding value of Log[M(HI)/M$_{*}$)] is -0.62. Presence of cold gas in Es has been interpreted as indicating a continuing growth of these galaxies up to the recent past \citep{Serra10,Thom12}. They may accrete cold gas directly from the field or via merger (major/minor). Although  SDSS J1229+0001 has the gas mass 4$\times$10$^{8}$ M$_{\sun}$ and gas mass fraction -0.62 the observed star-formation rate of 0.26 M$_{\sun}$/yr seems to an upper limit for similar mass (M$_{B}$  $\approx$ -16.5 mag) star-forming dwarf galaxies \citep[][Figure 6]{Lee09}. Assuming an HI consumption fraction of 10\%, the star-formation activity only last 160 Myr with current SFR. So it is very likely to be just a temporary phase.

\cite{Hallenbeck12} in a study of a  sample of gas bearing early-type dwarf galaxies in Virgo cluster propose  two possibilities  -- either such galaxies have recently accreted their gas or they are in the last stages of star-formation. In particular, for those which have detectable star-forming activity they suggest the later. Is the similar case for the SDSS J1229+0001? The observed relatively high emission line metallicity may be well explained according to the fundamental relation between gas-mass fraction  star-formation rate and emission line metallicity where where emission line metallicity anti-correlates with gas-mass fraction for given star-formation rate \citep{Lopez10}. We also find that two of \cite{Hallenbeck12} objects, only for which we can derive emission line metallicity from the SDSS spectroscopy, have similar value of 12+log(O/H) = $\sim$8.45. The study of \cite{Geha12} showed that field galaxies of dwarf regime always have some residual star formation. The SDSS J1229+0001 is located in relatively far form the massive halo, so the presence of star-formation is in line of Geha et al prediction.

How would it evolve after cease to form star in the centre? Any dynamical and stellar population evolution of compact star-forming object is likely to decrease the surface brightness  and expand the size \citep{Assmann13,Pfalzner13,Wellons15,Badry16}. It is certain that the central surface brightness in optical band decreases as it becomes older. Does it look similar to a typical low-surface brightness galaxy dE/dS0 or still remain low mass high surface brightness early-type galaxy like E/S0?  The mean surface brightness 17.53~mag~arcsec$^{-2}$ in H-band is $\sim$2 mag higher compared to average value of dE studied in \cite{Janz14}. Similarly, it is unusually compact having H-band half-light radius similar to NGC4486A. NGC4486A is a compact low mass-early type galaxy in Virgo cluster which also shows unusual burst of star-formation at the centre \citep{Prugniel11}. NGC4486A is $>$4.5 mag brighter than SDSS J1229+0001. In this regard, SDSS J1229+0001 maybe considered a low luminosity version of NGC4486A.

In summary, we find a couple of ignored morphology galaxy which maybe more similar to high z galaxies \cite[e.g.][]{Tadaki15}.  The red colour and relatively high emission line metallicity of these objects might suggest a deficiency gas mass fraction while actively forming stars at the end stage of evolution. Compact star-burst and dust-dominated galaxies are expected to be common in the early universe -- are these galaxies a low redshift counterpart?   In fact, the formation of compact ellipticals in the high-redshift universe has been associated with merger-induced star-burst activity with dissipative collapse of star-forming gases \citep{Kormendy92,Hopkins09}.  It is also important explore why there are no such compact, massive, galaxies (of both star-forming and non star-forming types) in the nearby universe. While more detailed comparative analysis of structural and stellar population properties of more complete sample would be necessary to establish an unambiguous evolutionary connection, these observational findings themselves maybe an important step toward toward this. In the next series of publication we aim to explore further.

\section*{Acknowledgments}
This study has made use of NASA's Astrophysics Data System Bibliographic Services and the NASA/IPAC Extragalactic Database (NED). SDSS data were queried from the SDSS archive. Funding for the SDSS/SDSS-III has been provided by the Alfred P. Sloan Foundation, the Participating Institutions, the National Science Foundation, the U.S. Department of Energy, the National Aeronautics and Space Administration, the Japanese Monbukagakusho, the Max Planck Society, and the Higher Education Funding Council for England. The SDSS Web Site is http://www.sdss.org.


\end{document}